\newcommand{\beq}{\begin{equation}}
\newcommand{\enq}{\end{equation}}
\newcommand{\beqarr}{\begin{eqnarray}}
\newcommand{\enqarr}{\end{eqnarray}}
\newcommand{\eqref}[1]{Eq.\ (\ref{#1})}
\newcommand{\figref}[1]{Fig.\ \ref{#1}}

\newcommand{\PRL}[1]{Phys.\ Rev.\ Lett. {\bf #1}}

\documentstyle[aps,prl,multicol,epsf]{revtex}
\begin{document}

\title{
Computational Complexity of
Determining the Barriers to Interface Motion
in Random Systems
}

\author{A. Alan Middleton}
\address{
Department of Physics, Syracuse University,
Syracuse, New York 13244
}

\date{July 16, 1998}

\maketitle

\widetext 

\begin{abstract}
The low-temperature
driven or thermally activated
motion of several condensed matter systems is often modeled by
the dynamics of interfaces (co-dimension-$1$ elastic manifolds)
subject to a random potential.
Two characteristic quantitative features of
the energy landscape of such a many-degree-of-freedom
system are the ground-state energy and the magnitude of
the energy barriers between given configurations.
While the numerical determination of the former can be
accomplished in time polynomial in the system size, it is shown
here that the problem of
determining the latter quantity is NP-complete.
Exact computation of barriers is
therefore (almost certainly) much more
difficult than determining the exact ground states of interfaces.
\end{abstract}


\begin{multicols}{2}
\narrowtext

\section{Introduction}

Numerical computation has been extensively used for confirming
scaling relations from
analytical work and for computing exponents in the statistical
mechanics of disordered systems, such as random magnetic systems
\cite{RandMag}.
In order to calculate long wavelength and low frequency behaviors,
and to precisely determine exponent values, one needs to study a
number of samples of large dimension.
The utility of numerical
techniques strongly depends on how the computational
demands, such as the number of operations needed by an algorithm,
scales with system size.
Direct simulation by Monte Carlo techniques often cannot be
used to determine the exact
ground state in large systems, for example,
due to the extremely slow relaxation times
typical of disordered systems \cite{PinningBarrier}.
In contrast,
the exact ground state can be found for many systems
\cite{HuseHenley,CombOpt,MiddletonEtAl,RiegerReview},
using combinatorial optimization
techniques, with a computational time that grows polynomially
with the volume of the system $V$ (in practice,
the computational time often scales roughly
as $V^{b}$, with $1 < b < 2$ \cite{MiddletonEtAl,RiegerReview}.)
More detailed information about the ``energy landscape'' of the
model, besides ground state energies,
is needed to determine dynamical behavior, which
is generally modeled by studying thermal activation over
barriers between low-energy states
\cite{PinningBarrier,MikheevEtAl}.
It is shown in this paper
that, for many models that might be of physical interest,
the numerical study of barrier
heights is in the set of NP-complete problems \cite{NPComplete}.

To define NP-complete problems, one first
defines NP (non-deterministic polynomial time) problems \cite{NPComplete}.
A problem which is NP is one for which a proposed solution
be verified
in an amount of time that grows no faster than a polynomial
in the size of the problem definition,
given a model of a computer as a
Turing machine \cite{hardNote}.
An NP-complete problem is then defined as an NP problem such that if
there existed an algorithm that produced a solution in polynomial
time (polynomial in the size of the definition of the problem),
such an algorithm could be adopted to
solve any NP problem in polynomial time.
The class of NP-complete problems includes
the traveling salesman problem \cite{NPComplete}
and ground states of spin glasses
on general graphs \cite{NPspinglass}.
It is generally believed that no polynomial-time
algorithm exists for NP-complete problems, though a proof of this
is an outstanding problem in computer science.
Showing that determining barrier heights
can be an NP-complete problem therefore strongly suggests that no
polynomial time algorithm can be found that exactly determines
barrier heights.

\section{Interface model}

To address the computational complexity of a problem, a discrete
formulation must be given and the size of
the problem must be defined.
We consider here the study of the configurations of a
self-avoiding, connected $D$-dimensional
interface
embedded in a bounded region $X \subset {\bf R}^{d}$,
$d = D + 1$,
such as might be used to approximate a domain wall in a magnet.
The interface
separates $X$ into two disjoint
sets.
A discretization can be made by approximating
the interface by a set of $D$-dimensional
polyhedra,
which are the faces of
$d$-dimensional cells that partition the region $X$.
One might consider the cells to represent actual atoms (with
the polyhedral faces being polygons for $d=3$ lattices)
or as elements of a
coarse-grained description, with cells
having a size of typical disorder-induced
distortions in the elastic interface \cite{IMLO}.
The size of the interface problem can then be defined as the number of
cells $n$ in the decomposition of $X$, assuming that there is
an $n$-independent bound on the number of faces for any cell,
so that
the number of discrete faces which make up the interface is then
bounded by a constant multiple of $n$ \cite{noteA}.
The dynamics of the interface is 
a sequence of simple moves which move the interface
through primitive cells, one at a time, while maintaining
the self-avoidance
and connectedness constraints.

For this model of an interface, it is natural to reformulate the
interface problem as an Ising model
on a graph $G_0=(V_0,E_0)$ given by the cellular decomposition of $X$.
The nodes of the graph $V_0$ are identified with the primitive
cells ($E_0$ will be defined for special cases, in the next
paragraph.)
Each node is assigned a spin
variable $s_i$, which takes on values $\pm 1$, with
a sign depending on which side
of the interface cell $i$, $1 \leq i \leq n$, belongs to.
In general, the energy of the interface could be any function
of the $\{s_i\}$ and could have a number of {\em independent} values
exponential in $n$.
Determining the ground state of the interface given by
such an energy function would likely
necessitate the computation of a number of energies exponential
in the volume.
We do not consider this case here, as it is not of
physical interest.

A more restricted, but natural case, is to represent the energy
of an interface as the sum over interactions between the spins.
That is, consider the energy function
\beq
E = -\sum_{i,j} J_{ij} s_i s_j,
\enq
where the $J_{ij}$ are edge weights; when $J_{ij} \ne 0$, an edge
$e=(i,j) \in E_0$ is included in $G_0$.
We consider here only the case of $J_{ij} \ge 0$, as the case where
$J_{ij} < 0$ for some bonds is the spin-glass problem,
for which determining
even the ground state is an NP-complete problem,
for graphs of physical
lattices in more than two dimensions \cite{NPspinglass}.
The energy is equal to twice
the total weight of the bonds which
intersect the interface, up to the additive constant
$-\sum_{ij} J_{ij}$.
If the energy is given by nearest neighbor interactions,
i.e., $J_{ij} \ne 0$ only for cells $i$ and
$j$ that share a face,
this energy corresponds to a nearest neighbor
random-bond Ising model (RBIM).

The ground state energy of the interface is then
defined as finding the minimum of $E$ over all
configurations $\{s_i\}$, possibly subject to boundary conditions that
define the two regions of space that are to be separated.
Such ground state energies and configurations give a great
deal of information about the system, including the response to
boundary condition changes and sensitivity to random changes in
the $J_{ij}$ \cite{BrayMoore}.
Such information for the RBIM and many other problems
can be found using polynomial-time algorithms
\cite{MiddletonEtAl,RiegerReview}.

A more subtle characterization of
the energy landscape
is the energetics of
extremal paths in configuration space.
A sequence of configurations related by simple moves (flips
of single spins adjacent to the
interface) are the paths of interest in configuration space.
The cost of a
path is defined
as the maximal value of the interface energy over all
intermediate configurations in the path.
The barrier between two given configurations can
then be defined as the minimal cost over
all allowed paths between the two configurations.

More precisely,
a path $Q$ of length $M$ is defined by an initial
configuration $\{s_i\}$ and a sequence of
locations for single spin flips,
$\{Q_1, Q_2, \ldots, Q_M\}$,
with $1 \leq Q_j \leq n$ for all $1 \leq j \leq M$.
Each spin flip corresponds
to moving a cell from one partition (side of the interface)
to the other, with the sequence constrained by the
interface self-avoidance and connectedness conditions.
The result of a partial sequence of
spin flips $P_k^Q = (Q_1, ... , Q_k)$,
$1 \leq k \leq M$,
operating on a spin
configuration $\{s_i\}$ is given by its action on individual spins
\beq
P_k^Q s_i = (-1)^{n_{ki}} s_i,
\enq
where $n_{ki}$ is the number of times
$Q_l = i$ for $1 \leq l \leq k$.
$P_0$ is the identity operation.
The value of the barrier
$B(S_1, S_2)$
separating two spin configurations $S_1 = \{s^1_i\}$
and $S_2 = \{s^2_i\}$ is
then given by
\beq
B(S_1, S_2) = \min_{\{Q | P_M^Q(S_1) = S_2\}}
	\max_{k=0,\ldots,M(Q)} E[P_k^Q\{s_i\}].
\enq

\section{Barrier for an arbitrary graph}

Though the ground state of the random-bond or random-field
Ising magnet for a
given realization of bonds can be computed in amount of time
bounded by a polynomial in the number of bonds and sites
\cite{HuseHenley,CombOpt,MiddletonEtAl,RiegerReview},
the problem of finding the barrier between two states will be
shown here to belong to the class of NP-complete problems,
for the motion of a self-avoiding interface.
There are no known polynomial-time algorithms for solving NP-complete
problems and it is generally believed that such problems 
require a time for solution that grows faster than
any polynomial in the problem size \cite{NPComplete}.
The barrier problem is therefore (almost certainly) in a computational
complexity class distinct from that of the ground-state problem.

The barrier problem can be shown to be NP-complete
for general graphs.
This is done by a reduction \cite{NPComplete}
of a register allocation problem,
which has been shown to be NP-complete \cite{minmaxregister},
to the barrier problem.
Loosely stated (see
\cite{minmaxregister} for more detail),
the register allocation problem is
that of determining the amount of memory a CPU needs to use to
store intermediate results in the
evaluation of an arithmetical expression.
The problem size is given
by the size of the arithmetical expression, which is defined by a
set of parenthesized binary
operations.  Initially, no register memory is used,
but values for the variables in the expression are loaded
into registers and intermediate
results are stored
in registers, until
the expression is completely evaluated.
The cost to be minimized is the
maximum number registers in use
at any time during the computation.
The cost is affected by the order in which the binary operations
are evaluated and in which the values are loaded into registers.
The correspondence between the register allocation problem and
finding the minimal barrier to interface motion can be made
by identifying cells as the numerical values (original inputs to
the expression and the results of binary operations) needed
in the evaluation of the arithmetic expression and identifying
shared faces of the cells with the dependencies of results on
subexpressions.  The interface separates evaluated
numbers from intermediate results yet to be computed.
The energy of the
interface is taken to
be the number of cells that
contact one side of the interface; this is the number of intermediate
results that are available in registers for further evaluation of the
expression.
The optimization of register use is then equivalent to
finding the minimum energy barrier to moving
an interface through the graph determined by the
algebraic expression.
As the register allocation problem is NP-complete, the corresponding
(high-dimensional) interface problem is NP-complete.
I do not give a more detailed proof here, however,
as the following section shows that a restricted interface
barrier problem is NP-complete.
This directly implies the NP-completeness of the more general problem.
It is of interest, however, to note the close relationship between
the determination
of barriers in a type of interface motion and computational
resource problems
such as register allocation.

\section{Barrier for a loop in $d=2$}

Consider now the more specific
model of an interface described by a self-avoiding
closed path in a planar graph $G$.
The graph $G$ is dual to a graph $G_0$ which defines a
nearest-neighbor RBIM.
The edges of $G$ will be the faces of the cells defining interface
motion.
The face separating cells $i$ and $j$ is assigned
weight $2J_{ij}$.  Here we will restrict the $J_{ij}$ to
have non-negative half-integer values.
The energy of the loop is then the sum of the
weights of the edges it passes through.
This interface might define the surface of a magnetic domain in
two dimensions, for example.
Finding the barrier
that separates two configurations of a loop in the plane
will be referred to
here as the loop-barrier problem.
Note that the loop is permitted to vary in length.

The ground state of a loop in a plane can be determined in
polynomial time using maxflow algorithms
\cite{CombOpt,MiddletonEtAl,generalmincut}
on the graph $G_0$.
One can either find the
the minimum cost interface separating two points or regions,
for example, or a globally minimal
interface \cite{noteB}.
Not all loop or path problems in the plane are so easily
solved.
There is at least one example where finding the ground state
for a loop is an NP-complete problem:
the NP-complete problem of finding a Hamiltonian path
in a planar cubic graph
can be reduced to determining
the ground state of a loop in the
plane with {\em fixed
length} \cite{Machta}.
In contrast, finding the minimal {\em directed} path of fixed length
is easily shown
to be linear in the volume of the lattice, as shortest-path
algorithms can be used \cite{HuseHenley}.

It will be shown here that the loop-barrier problem is NP-complete by a
reduction of the problem planar 3-satisfiability (P3SAT)
\cite{NPComplete,Lich}.
That is, any algorithm which solves
the loop-barrier problem can be applied to
solve P3SAT problems and the time taken to translate 
an instance (realization) of the P3SAT problem
to a loop-barrier instance is bounded by a polynomial in the size of
the P3SAT instance.
Since P3SAT is NP-complete, it follows that determining the barriers
for loops in the plane (and, trivially,
barriers to the motion of closed surfaces in higher dimensions) is
NP-complete.

To define P3SAT, it is convenient to consider first the more general
problem of 3-satisfiability (3SAT) \cite{NPComplete}.
An instance of 3SAT is defined by a set of $p'$ Boolean variables
$B = \left\{ u_1, \ldots, u_{p'} \right\}$ and $q'$ clauses
$C = \left\{ c_1, \ldots, c_{q'} \right\}$.
A clause $c_i$ is a
triplet of literals,
\beq
c_i = \left\{ z^1_i, z^2_i, z^3_i\right\},
\enq
where each literal $z^a_i$ is either $u_j$ or
$\overline{u}_j$ (the negation of $u_j$)
for some $1 \leq j(a,i) \leq p'$.
The set of clauses
is said to be satisfiable if,
for some set of
truth assignments for the $\{u_j\}$,
at least one of the literals in $c_i$ is true,
for all $1 \leq i \leq q'$.
That is, there is a choice of values for the $\{u_j\}$ such that
the Boolean expression
\beq
(z_1^1 \vee z_1^2 \vee z_1^3) \wedge
(z_2^1 \vee z_1^2 \vee z_1^3) \wedge
\ldots \wedge (z_{q'}^1 \vee z_{q'}^2 \vee z_{q'}^3)
\label{eqnSAT}
\enq
is true.
The problem 3SAT is to determine whether \eqref{eqnSAT} is
satisfiable.

A given satisfiability problem can be identified
with a graph $G'=(V',E')$,
with vertices 
\beq
V = \left\{
	v_k : 1 \leq k \leq p'+q'\right\} = B \cup C,
\enq
(renaming variables $\{u_1,\ldots,u_{p'}\}$ and clauses
$\{c_1,\ldots,c_{q'}\}$ as $\{v_1,\ldots,v_{p'}\}$
and
\linebreak
$\{v_{p'+1},\ldots,v_{p'+q'}\}$, respectively.)
The edge set $E'$ is defined by
\begin{eqnarray}
E' & = & \{
	(v_i, v_j) :
	1 \leq i \leq p',\,
	p' < j \leq p'+q',\, \\
& &\ \ \ 	u_i \in c_{j-p'}\ {\rm or}\ \overline{u}_j \in c_{j-p'}
	\}.\nonumber
\end{eqnarray}
The instances of P3SAT are just those 3SAT instances
whose graphs $G'$ can be embedded in a plane (the vertices and edges
can be placed so that edges do not intersect.)

\begin{figure}
\begin{center}\leavevmode
\epsfxsize=6.5cm
\epsffile{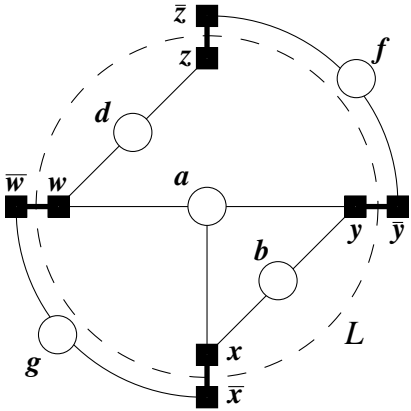}
\end{center}
\caption{
A particular realization of P3SAT (planar 3-satisfiability),
represented
as a planar graph $G$.
The ``splitting'' of the Boolean variables into a variable and its
negation are shown as pairs of literals (filled squares)
connected by thick edges.
Clauses are indicated by open circles.
Edges between literals and clauses are indicated by thin lines.
Clauses have edges to a maximum of three literals and literals
have edges to a maximum of two clauses.
The loop $L$ passes between
all pairs of literals, but does not intersect any of the edges between
variables and clauses.
The particular Boolean expression represented
by this graph is
$a \wedge b \wedge d \wedge f \wedge g =
(x \vee y \vee w) \wedge (x \vee y) \wedge (w \vee z) \wedge
(\overline{y} \vee \overline {z}) \wedge
(\overline{w} \vee \overline {x})$.
Determining whether such a planar
graph is satisfiable, that is, whether there
exists an assignment of truth values to the Boolean variables for
which the expression is true, is an NP-complete problem.
}
\label{figP3SAT}
\end{figure}

P3SAT has been shown to be NP-complete \cite{Lich}.
In the proof developed
in Ref. \cite{Lich}, it is shown that any 3SAT
instance can be polynomially reduced to a P3SAT instance.
The construction of the P3SAT graph $G$
($p$ variables and $q$ clauses)
from $G'$
($p'$ variables and $q'$ clauses)
has three properties
which are central to the following reduction:
{\em Property 1 ---} There exists
a cyclic path $L$
which intersects none of the edges of the graph $G$, but
passes through all of the vertices $v_i$ with $1 \leq i \leq p$,
that is, all
of the nodes corresponding to the variables $\{u_i\}$.
{\em Property 2 ---} The edges between variables and clauses
can be arranged so that only variables (negated variables) are in
interior (exterior) clauses.
More precisely,
define the two sets $C_{\rm int}$ and $C_{\rm ext}$, consisting of the
clauses interior and exterior to the cycle $L$.
Then, given any variable index $1 \leq i < p$,
for all $c \in C_{\rm ext}$, $u_i \not\in c$, and
for all $c \in C_{\rm int}$, $\overline{u}_i \not\in c$ \cite{noteD}.
In particular, variables $u_j$ may be ``split'' into two nodes
of the graph, representing the literals $u_j$ and $\overline{u}_j$,
so that a variable and its complement are separated by $L$.
{\em Property 3 ---} No literal is a member of more than two
clauses.
Properties (2) and (3) are a restatement
of Lemma 1 in Ref.~\cite{Lich}.
An example of a P3SAT problem and its corresponding graph, showing
the path $L$ through
``split'' variables, is
shown in \figref{figP3SAT}.

A loop-barrier problem can be constructed from a P3SAT instance with
Properties (1-3),
with the number of steps in the construction polynomial in the
size of the P3SAT instance, such that a computation of the barrier
for loop motion
would determine the satisfiability of the P3SAT instance.
The loop considered is a self-avoiding (no vertex is shared by more than
two edges) returning walk on a lattice (i.e., a simple cycle.)
In order to make the closest connection to problems in condensed matter
physics, the problem is mapped onto a regular lattice, though a
slightly simpler reduction onto
a general planar graph can be made.
The energy of a loop is defined by summing edge weights
$E_{ij} = 2J_{ij}$ for the ``faces'' (i.e., edges) separating
nearest neighbor pairs of cells $(ij)$.
As in the general case discussed above,
the allowed moves are those which change the path by moving the
loop sequentially over
single cells (a polygon in this planar
case.)
The barrier problem is to determine the barrier between two loop
configurations, given the elementary cell-crossing moves and
the constraint of self-avoidance.

The central notion behind the reduction can be
summarized briefly.
The barrier problem will be constructed so that
the energy of a loop, initially coinciding with the cycle
$L$ (\figref{figP3SAT}), will be lowered by distorting it so that
it passes through locations corresponding to the clauses.
Each such distortion can lower the loop energy by no more than a unit
amount.
These distortions of the loop will implicitly require
a choice of truth values for
the variables, by the direction in which the loop is distorted.
If all of the clauses can be satisfied, by choosing truth
values for a sufficient number of variables, the energy
of the loop will be
lowered from its initial value by an amount $q$, allowing
the loop to move into an
otherwise forbidden region, which has a barrier $q$ to enter.
Hence, if the P3SAT instance is
satisfiable, the barrier to motion of the loop into the forbidden
region
will be zero.
If the P3SAT instance is not satisfiable, the barrier
to motion into the forbidden region will be positive.
The satisfiability
of the instance therefore holds if and only if the barrier is zero.

\begin{figure}
\begin{center}\leavevmode
\epsfxsize=6.5cm
\epsffile{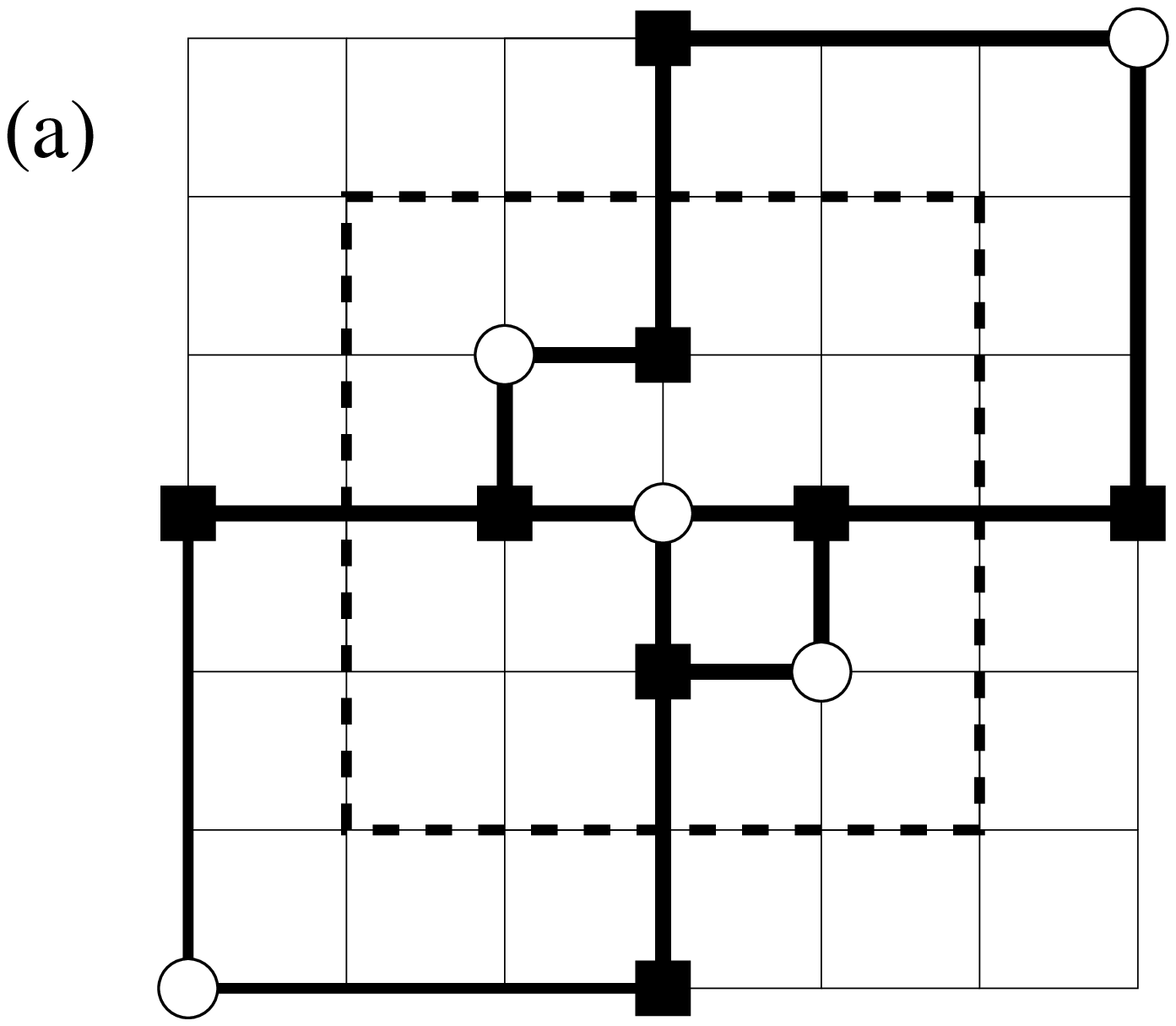}
\end{center}
\begin{center}\leavevmode
\epsfxsize=6.5cm
\epsffile{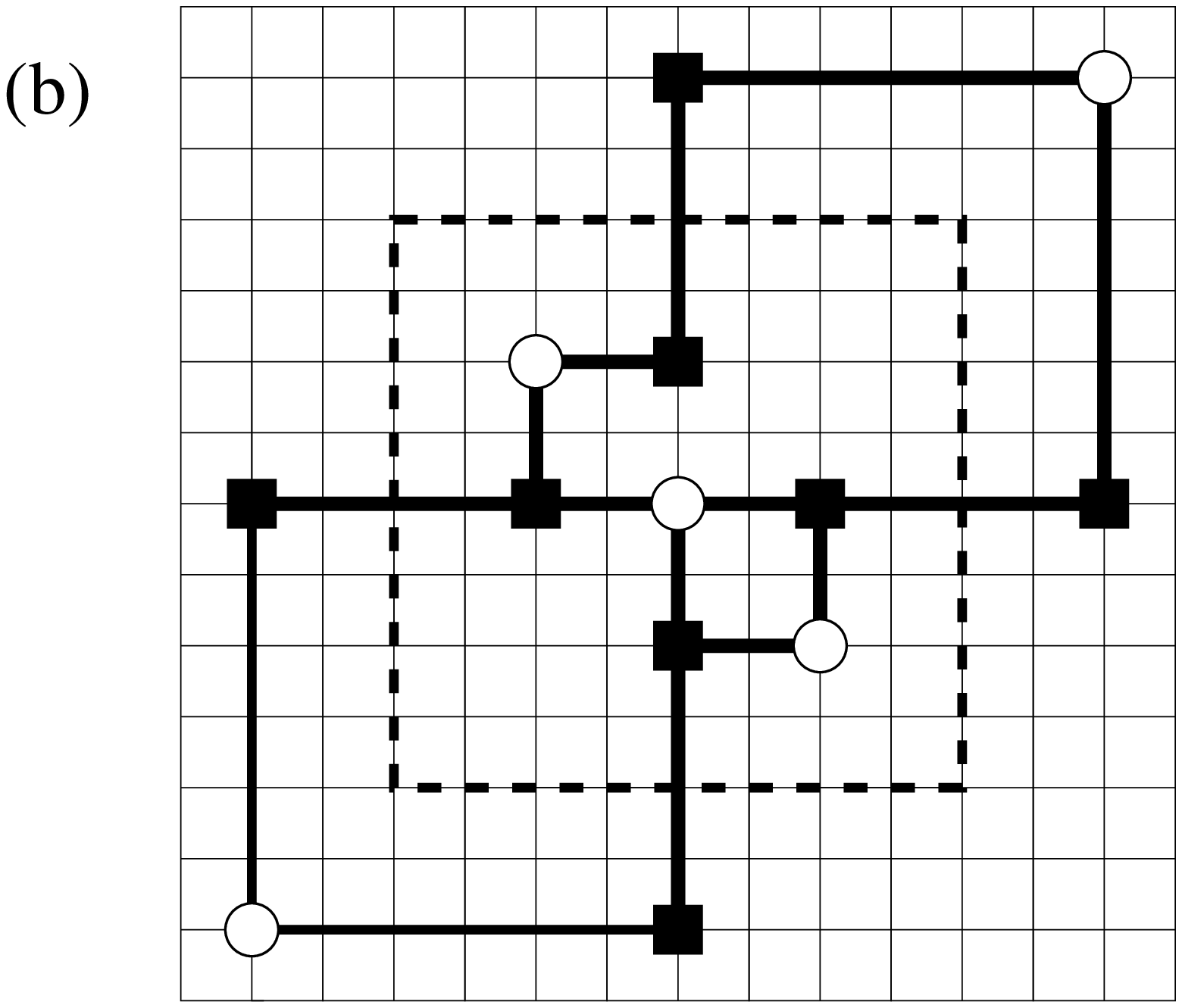}
\end{center}
\begin{center}\leavevmode
\epsfxsize=6.5cm
\epsffile{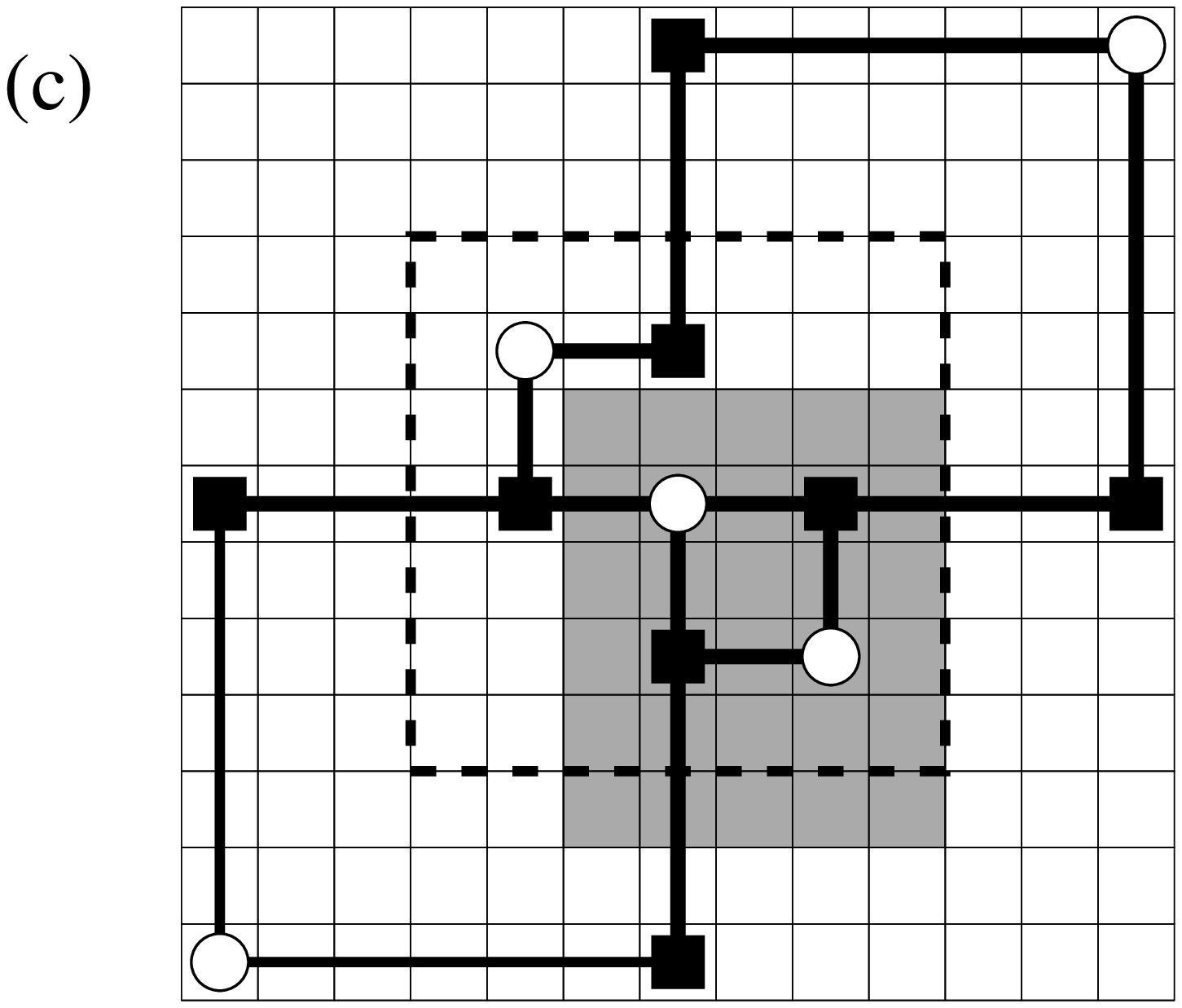}
\end{center}
\caption{
(a) The embedding of the 
graph $G$ of \figref{figP3SAT} into a square lattice $A$;
the loop $L$ is indicated by the dashed line and edges in $G$
are indicated by thick lines.
(b) An illustration of
the refinement $A_2$ of $A$ constructed by halving edge lengths.
(c)
The square lattice dual to $A_2$ (excluding the point corresponding
to the boundary cycle), $A_d$, for the sample graph, with the loop $L$
contracted so that its edges lie in the dual graph.  The nodes
of $G$
now correspond to squares while the edges of $G$
correspond to sequences of squares.
The shaded region indicates the area illustrated in detail
in \figref{figFinal}.
}
\label{figEmbed}
\end{figure}

The first step of the construction is to embed the P3SAT graph $G$,
with the loop $L$ guaranteed by Property (1)
and with variables split into literals,
into a square grid $A$ of size polynomial
in $p+q$  \cite{Valiant} (\figref{figEmbed}(a).)
The grid $A$ is then refined by dividing each square into four
smaller squares and then adding a single border layer of squares
around the
whole lattice, resulting in the
grid $A_2$ (\figref{figEmbed}(b).)
This step insures that all nodes corresponding to nodes
of $G$ are separated
from each other and from the edges of
the loop $L$ by a distance of at least two lattice spacings, and that
no node is on the boundary of the grid.
Given this refined embedding of $G$, the grid $A_d$ is defined as
the dual of
$A_2$, excluding the dual node that corresponds to the
boundary cycle of $A_2$.
Each node of the lattice embedding of G will now correspond to a
square plaquette in $A_d$ (\figref{figEmbed}(c).)
The final lattice is constructed by
drawing a diagonal with uniform direction
across the plaquettes of $A_d$, defining
a triangular grid $T$,
with each node of $G$ corresponding to two
triangles.
The cells of $T$ will be the cells defining interface motion.

\begin{figure}
\begin{center}\leavevmode
\epsfxsize=6.5cm
\epsffile{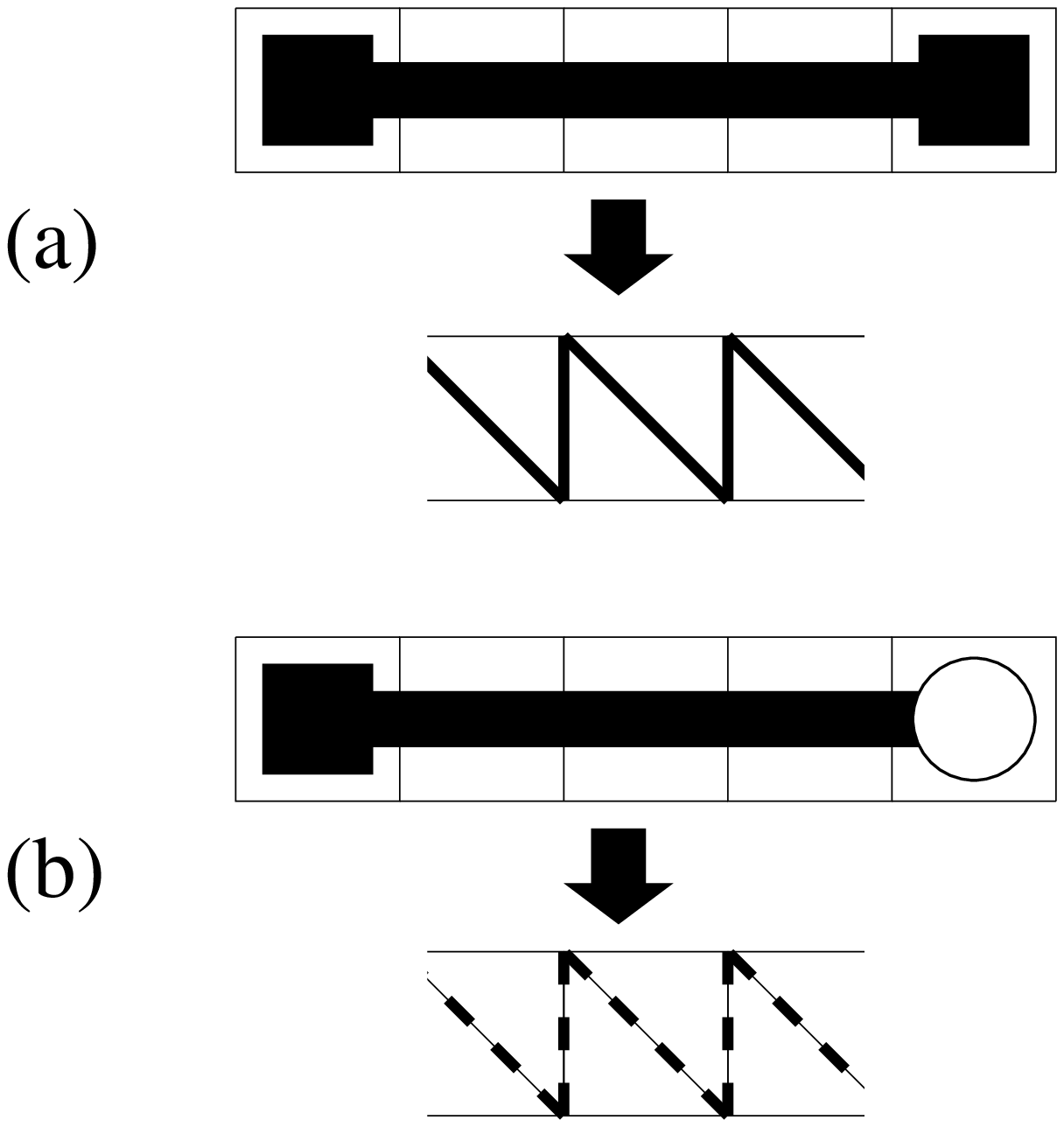}
\end{center}
\caption{
Pictorial representation of rules for assignment of weights in
the graph $T$.
Heavy edges have weight $2$, dashed edges have weight $1$, and
thin edges have weight $0$.
(a) Edges that intersect an edge in $G$ between literals
are assigned a weight of $2$.
(b) Edges that intersect an edge in $G$ between a literal
and a clause are assigned weight $1$.
}
\label{figWeights}
\end{figure}

\begin{figure}
\begin{center}\leavevmode
\epsfxsize=6.5cm
\epsffile{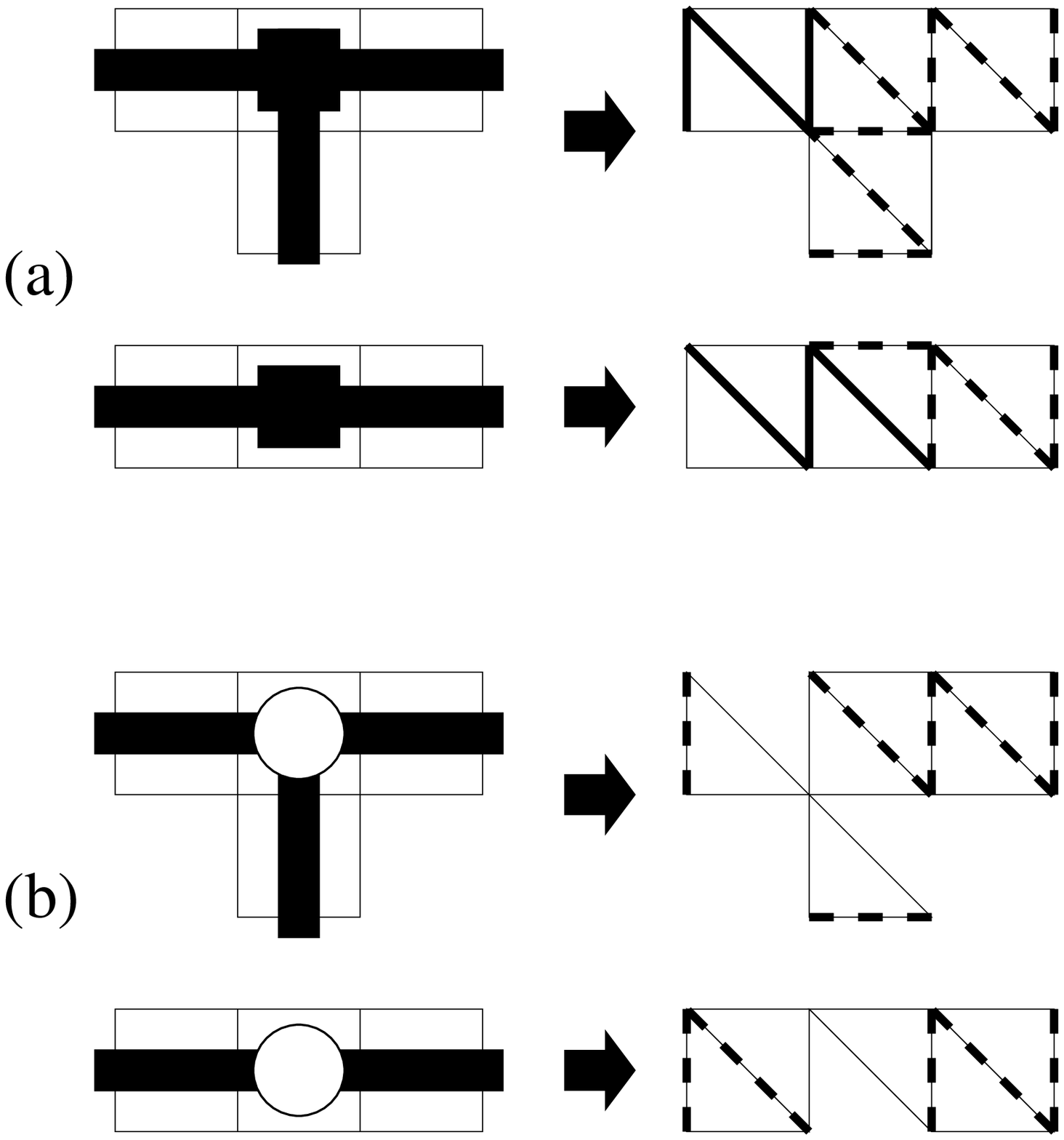}
\end{center}
\caption{
Pictorial representation of rules for assignment of weights in
the graph $T$, for edges near literals and clauses.
(a) The two triangles corresponding to a literal that is a member
of two clauses are designed
to share the weight between the two paths to the clauses.
In this figure, the left branch connects to the negation
of the literal, while
the down and right branches connect to a clause.
If a literal is a member of only one clause, the literal
``absorbs'' one unit of weight.
The left side of the literal is connected to another literal, while
the right branch is connected to a clause, in this example.
(b) Clause
regions are designed to lower the weight of the loop,
when it intersects
one corner of the clause square, by a unit amount.  Triangles that are
at the ends of edges incident upon a clause meet at a single node
of the square corresponding to a clause.  This node is referred
to as a clause node. The self-avoidance
constraint prohibits the loop from moving ``through''
the clause nodes or
the loop from simultaneously intersecting the clause node from
more than one direction and thereby lowering the loop energy more
than once per clause.
}
\label{figJunctions}
\end{figure}

\begin{figure}
\begin{center}\leavevmode
\epsfxsize=6.5cm
\epsffile{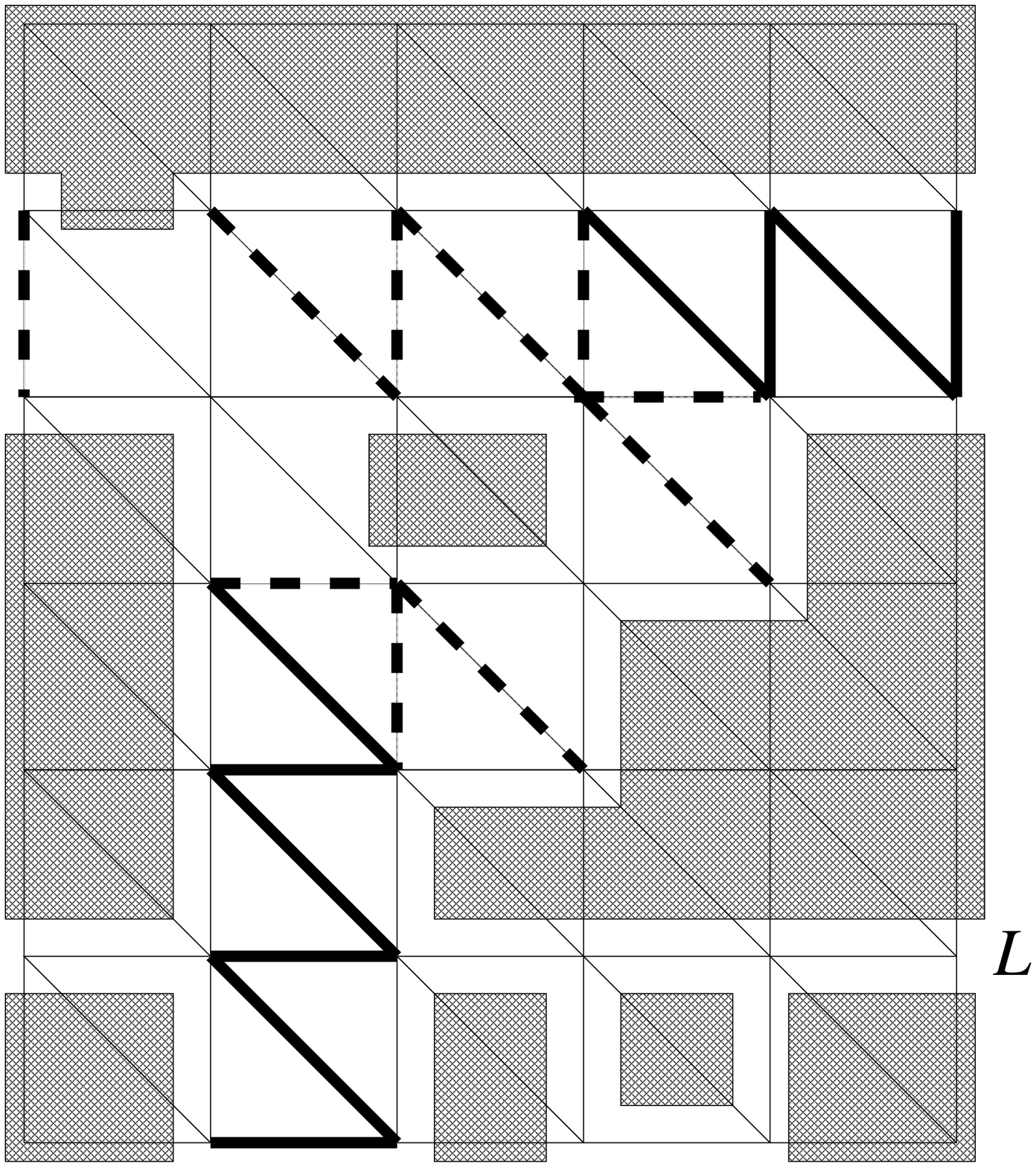}
\end{center}
\caption{
In the final step of the reduction, the square plaquettes of
$A_2$ are divided into two triangles by the introduction of
diagonal edges.  Weights are then assigned to the edges, according
to the rules listed in the text and illustrated in \figref{figWeights}
and \figref{figJunctions}.
Edge
weights for the shaded region of \figref{figEmbed}(c) are shown here.
Edges which intersect
the hatched areas are assigned a weight of
$q$. The weight of edges on the loop $L$ that are not used to 
connect literals is set to zero.
The heavy line segments indicate edges of weight $2$ and
the dashed segments indicate edges of weight $1$.  Thin edges
are assigned a weight of $0$.
The final part of the assignment of the weights for $T$
is to set the weights for three edges near $L$ to zero,
as indicated by the unhatched edges at the bottom right of the
figure.
The barrier problem is to distort $L$ with minimal maximum
cost so that it
passes through these three zero-weight edges.
}
\label{figFinal}
\end{figure}

The correspondence between the P3SAT instance
and the edge costs is best
made clear by reference to 
\figref{figWeights} and \figref{figJunctions},
which illustrate weight assignment
rules, and \figref{figFinal}, which
illustrates a subgraph of $T$ for the sample expression.
First, weights are assigned to edges of $T$ that
intersect the edges of $G$ (\figref{figWeights}.)
The edges of $T$ that intersect
edges of $G$ between literals are assigned a weight of $2$.
The edges of $T$ that intersect edges between literals
and clauses are assigned weight $1$.
Next, the weights of edges in $T$ that are incident upon
the edges corresponding
to squares in $A_d$ that contain a literal or a clause
are assigned (\figref{figJunctions}.)
The weight of the diagonal edge corresponding to a literal
is set to $2$.
The pair of triangles
corresponding to the node for the literal then divides the weight into
two edges of unit cost, if the literal is a member of two clauses.
If a literal is a member of only one clause, one of the edges of the
square that does not intersect an edge of $G$ is set to have a weight
of $1$, so that the literal will ``absorb'' one unit of cost and one
unit of energy can leave the square.
The most complicated part of the construction is the assignment of
weights near clauses (\figref{figJunctions}(b)).
Note that, near clauses, these rules may override the rule
shown in \figref{figWeights}(b).
Up to three edges of $G$ may enter a clause square in $A_d$.
One corner of the clause square is chosen to represent the clause.
This corner will be referred to as a clause node.
The weights in $T$
near this node are set so that (a) a primitive move that
places the interface on this corner reduces the weight of the loop
by a unit amount and (b) the interface cannot move ``forward'' through
the node at low cost.
These requirements (a) and (b) are met by
identifying triangles containing the clause node with
the ends of the edges of $G$ between literals and clauses.
These triangles have two edges of weight
zero and the other
edge with weight one, with the zero weight edges incident upon the
corner representing the clause.
Note that due to the self-avoidance
constraint, a loop may pass through a given clause node at most once.
The horizontal and vertical
edges that border the edges between
clauses and variables, and between literals, and have not yet been
assigned a weight, are set to have weight 0.
The weight of edges on $L$ that do not intersect edges of $G$ are
set to zero.
The remainder of the edges
(edges completely covered by the hatched regions
in \figref{figFinal}) are set to have weight $q$.
To complete the definition, the weights of three 
edges near $L$, which would otherwise be $q$ by the previous step,
are set to zero, as shown in \figref{figFinal}.

Now that the graph and
edge weights for a loop-barrier problem
have been defined, the initial and final
configurations can be described.
The initial path is taken to be the path describing the loop $L$.
The total weight of $L$ is $2p$.
The final path is taken to be the one differing from $L$ by only
two primitive moves and that
passes through the three edges set to zero weight in the final step
of the weight definitions.
The final configuration is also of weight $2p$.
The barrier problem is to determine whether there is a barrier of
{\em zero} energy separating the initial and final path.
This will be so if and only if
the initial path can be distorted so that
its energy is first lowered by $q$ from the initial energy.
This allows a bond coinciding with $L$
to be moved onto the high cost diagonal near the zero weight
goal edges without
exceeding the initial energy.
Note that while the loop cost remains greater than $2p-q$,
all moves that do not raise the
loop cost above $2p$ are neutral in cost, except for those that
move the loop onto a clause node and thereby
lower the energy by a unit
amount.

In order to lower the loop energy to $2p-q$,
{\em the initial loop $L$ must be distorted until it
intersects all of the clause nodes;
such a path configuration can be reached without
exceeding the initial energy if and only if
the given instance of P3SAT can be satisfied.}
First, suppose that such a path configuration can be reached.
Property (2) implies that moving
the loop $L$ in a particular direction
(neglecting high cost moves)
chooses a truth value for a variable by moving
the loop through the region
corresponding to a literal;
the loop cannot be in a state where it has
passed through both literals for a single
variable an odd number of times,
without first passing through a barrier of at least
$q$ in energy.
Once a literal is chosen, all of the
unsatisfied clauses which it belongs to
can be satisfied by distorting the loop onto the corresponding
clause node.
Given a loop which intersects
all clauses, values for the variables which satisfy
the expression can thus be directly deduced from which literals
the loop has passed through an odd number of times.
If a loop configuration of weight $2p-q$
reachable from $L$ without raising
the energy above $2p$ exists, the P3SAT instance can be satisfied.
Conversely, suppose that the P3SAT instance can be satisfied.
There must then exist
an assignment of truth values to the variables which
satisfy the instance.
Given such an assignment,
the loop $L$ can be moved so that it passes
over the corresponding literals with zero cost.
The loop can then be moved to simultaneously
pass through all $q$ clause nodes without raising
the loop cost at any time, since, by the
satisfiability of the P3SAT instance, each clause has
as a member one of the literals over which the loop has been
moved.
The loop will then have a weight $2p-q$.

Given a
loop configuration of weight $2p-q$
reachable without moves of cost $q$,
the loop can then be moved onto
the high cost diagonal
near the goal edges, raising its energy by $q$ back to $2p$.
The loop energy can then immediately
lowered by the same amount, by moving the loop onto
the zero weight goal edges.
The earlier moves that lowered the energy by $q$ can
then all be reversed, returning the remainder
of the loop to its original state.
The height of the barrier between initial and final
loop configurations is then zero.
This sequence of moves exists if and only if a loop configuration
of weight $2p-q$ can be reached with zero barrier.
Answering the question about the magnitude of
the loop-barrier is therefore equivalent to determining
whether the 
assignment
instance from which it was derived can be satisfied.
Since P3SAT is NP-complete, the general problem of
determining the barrier to motion of a loop in a plane is also
NP-complete.

\section{Comments}

In a general sense,
physical barrier problems can be related to resource allocation
problems which are NP-complete.
It has been shown here that the problem of determining the exact
barrier to a self-avoiding loop in the plane, given integer
edge weights bounded by the volume of the system ($q < n$,
where $n$ is the number of triangles in $T$), is NP-complete.
Physically, this is most closely related to the barrier to
the motion of a self-avoiding interface in an RBIM.
The proof of this result is based upon a mapping between
the assignment of
truth values that satisfy a Boolean expression
and
the distortions of a loop necessary to lower its energy by a
given amount.
The loop energy
can be lowered sufficiently to cross a barrier towards the goal
loop if and only if the Boolean expression can be satisfied.
An immediate application of this result
is to undirected paths in samples with periodic
boundary conditions.
A region interior to the loop $L$ that does not intersect any
of the edges corresponding to $G$ or the final loop
configuration can be
removed from $T$, so that
the loop-barrier problem corresponds to the motion of a loop on
an annulus.
This implies that the barrier problem for a 
periodic path in $1+1$-dimensions is NP-complete, in the general
case of a
self-avoiding cyclic path \cite{noteE}.
It is unclear at this time whether determining the
barrier to motions of a {\em directed} path \cite{MikheevEtAl}
in $1+1$-dimensions is an NP-complete problem.
Even if such a problem is NP-complete,
this does not rule out the existence of heuristic methods which
can give useful upper and lower bounds on the barrier, such as
those described in Ref.~\cite{MikheevEtAl}.

This work has been supported by the National Science Foundation
under grant DMR-9702242 and a 
Fellowship from the Alfred P. Sloan Foundation.


\end{multicols}

\begin{references}
\bibitem{RandMag}
For reviews, see, for example, 
K. Binder and A. P. Young, Rev. Mod. Phys. {\bf 58}, 801 (1986);
T. Natterman and P. Rujan, Int. Jour. Mod. Phys. B {\bf 3},
1597 (1989);
{\em Heidelberg Colloquium on Glassy Dynamics}, ed.
J. L. van Hemmen and
I. Morgenstern (springer-Verlag, Heidelberg, 1987);
H. Rieger, in
{\em Annual Reviews of Computational Physics II}, ed. D. Stauffer, 
(World Scientific, Singapore (1995), p. 295. 
H. Rieger, in {\em Annual Review of Computational Physics II}, (World
Scientific, Singapore, 1995).
\bibitem{PinningBarrier}
J. Villain, Phys. Rev. Lett., {\bf 52}, 1543 (1984);
G. Grinstein and J. F. Fernandez, Phys. Rev. B {\bf 29}, 6398 (1984);
M. V. Feigel'man, V. B. Geshkenbein, A. I. Larkin, and V. M. Vinokur,
Phys. Rev. Lett. {\bf 63},
2303 (1989). 
\bibitem{HuseHenley}
D. A. Huse, C. L. Henley, \PRL{54}, 2708 (1985).
\bibitem{CombOpt}
J. C. Angle d'Auriac, M. Preissmann, and R. Rammal,
J. Phys. Lett. {\bf 46}, L-173 (1985);
A. T. Ogielski, \PRL{57}, 1251 (1986).
\bibitem{MiddletonEtAl}
A. A. Middleton,
Phys. Rev. E {\bf 52}, 3337 (1995); 
C. Zeng, A. A. Middleton, and Y. Shapir,
Phys. Rev. Lett. {\bf 77}, 3204 (1996), cond-mat/9609029.
\bibitem{RiegerReview}
H. Rieger, in {\em Advances in Computer Simulation},
ed. J. Kertesz and I. Kondor,
(Springer-Verlag, Heidelberg, 1998), cond-mat/9705010.
\bibitem{MikheevEtAl}
L. V. Mikheev, B. Drossel, M. Kardar,
Phys. Rev. Lett. 75, 1170 (1995), cond-mat/9503020.
\bibitem{NPComplete}
For an introduction to computational complexity,
see M. R. Garey and D. S. Johnson, {\em Computers and Intractability}
(W. H. Freeman and Company, New York, 1979);
C. H. Papadimitriou, {\em Computational Complexity}
(Addison-Wesley, 1994).
\bibitem{hardNote}
NP-completeness is defined using ``decision problems'',
which require a yes$/$no answer to a question, given
the problem data.  Numerical questions, such
as determining the ground-state energy, can be reformulated as
questions about bounds (on the energy) in order
to translate the problem into a decision problem \cite{NPComplete}.
\bibitem{NPspinglass}
F. Barahona, J. Phys. A {\bf 15}, 3241 (1982).
\bibitem{IMLO}
A. I. Larkin and Yu. N. Ovchinnikov, J. Low Temp. Phys. {\bf 34},
409 (1979);
Y. Imry and S.-K. Ma, Phys. Rev. Lett. {\bf 35}, 1399 (1975).
\bibitem{noteA}
Barrier heights to interface motion can be computed exactly in
polynomial time if the interface size scales no faster than a power
of the logarithm of $n$, which could be the case for highly anisotropic
decompositions of $X$.
\bibitem{BrayMoore} A. J. Bray and M. A. Moore,
Phys. Rev. Lett. {\bf 58}, 57 (1987).
\bibitem{minmaxregister}
R. Sethi, SIAM J. Comput. {\bf 4}, 226 (1975).
\bibitem{generalmincut}
For discussion of minimum-cut and maximum-flow problems, see, for
example,
{\em Introduction To Algorithms},
T. H. Cormen, C. E. Leiserson, and R. L. Rivest
(MIT Press,
Cambridge, Massachusetts, 1990).
\bibitem{noteB}
For problems of physical
interest with a sufficient density
of non-zero bonds, it is likely that the
minimum cost interface will cut only a very
few bonds, as
loop costs will generally grow with loop size,
in a graph with a percolating density of
non-zero edge costs (also see \cite{Machta}.)
\bibitem{Machta}
J. Machta, J. Phys. A {\bf 25}, 521 (1992).
\bibitem{Lich}
D. Lichtenstein, SIAM J. Comput. {\bf 11}, 329 (1982).
\bibitem{noteD}
This is a slight modification of Lemma 1 in Ref.~\cite{Lich}.
As a matter of convention, the variables on
the loop can be redefined by negating their meaning
where necessary, so that
all of the clauses with negations of variables are exterior to $L$.
\bibitem{Valiant}
L. G. Valiant, IEEE Trans. Comput. {\bf C-30}, 135 (1976).
\bibitem{noteE}
I would like to thank S. Coppersmith and D. McNamara for discussions
on this point.
\end{references}
\end{document}